\documentclass[10pt,aps,pre,amsmath,amsfonts,amssymb,floatfix,superscriptaddress,notitlepage,nofootinbib,twocolumn]{revtex4-1}
\usepackage{mathrsfs} \usepackage{graphicx,color} 
\usepackage{multirow}
\usepackage{bm}

 \let\b=\beta \let\g=\gamma 
 \let\z=\zeta  
    
\let\s=\sigma

\def\CC{{\cal C}}

\def\DD{{\cal D}} 
  
\def\ZZ{{\cal Z}}

\def\de{\mathrm d}

\def\to{\rightarrow}

\newcommand{\beq}{\begin{equation}} \newcommand{\eeq}{\end{equation}}

\begin{document}
\title{
High dimensional optimization under non-convex excluded volume constraints}

\author{Antonio Sclocchi}
\affiliation{Institute of Physics, \'Ecole Polytechnique F\'ed\'erale de Lausanne, 1015 Lausanne, Switzerland}

\author{Pierfrancesco Urbani}
\affiliation{
Institut de physique th\'eorique, Universit\'e Paris Saclay, CNRS, CEA, F-91191 Gif-sur-Yvette, France
}

\begin{abstract}
We consider high dimensional random optimization problems where the dynamical variables are subjected to non-convex excluded volume constraints. We focus on the case in which the cost function is a simple quadratic cost and the excluded volume constraints are modeled by a perceptron constraint satisfaction problem.
We show that depending on the density of constraints, one can have different situations. If the number of constraints is small, one typically has a phase where the ground state of the cost function is unique and sits on the boundary of the island of configurations allowed by the constraints. In this case, there is a hypostatic number of marginally satisfied constraints. If the number of constraints is increased one enters a glassy phase where the cost function has many local minima sitting again on the boundary of the regions of allowed configurations. At the phase transition point, the total number of marginally satisfied constraints becomes equal to the number of degrees of freedom in the problem and therefore we say that these minima are isostatic. We conjecture that by increasing further the constraints the system stays isostatic up to the point where the volume of available phase space shrinks to zero.
We derive our results using the replica method and we also analyze a dynamical algorithm, the Karush-Kuhn-Tucker algorithm, through dynamical mean-field theory and we show how to recover the results of the replica approach in the replica symmetric phase.
\end{abstract}

\maketitle

\section{Introduction}
Several complex systems can be described by a large number of variables whose dynamics is trying to optimize some cost function under a set of constraints. 
In the most general setting, the cost function is arbitrary and the constraints may be either inequalities or equalities. 
In this work, we consider the generic situation in which there is a set of $N$ real variables $\underline x = \{x_1,\ldots,x_N\}$ and a set of $M=\alpha N$ inequality constraints
\beq
\CC=\{g^\mu(\underline x) \geq 0, \mu=1,\ldots,M=\alpha N\}\:.
\label{problem_generic}
\eeq
each characterized by a function $g^\mu(\underline x)$ and we consider a cost function $H(\underline x)$.

An example of this situation is found in classification problems in machine learning.
In the simplest setting of binary classification, one has a set of inequalities that enforce the correct separation of the training set into two classes. The parameters of the machine (that correspond to the variables $x_i$ here) must be adjusted to correctly satisfy these constraints. Therefore, the supervised learning task is recast into a continuous constraint satisfaction problem (CCSP). 
In many cases one wants to enforce that, once a solution of the problem is found, it satisfies some requirements. For example, in order to prevent numerical instabilities, one would like to have the parameters of the network be quantities that are of the good order of magnitude and do not explode to infinity. To achieve this, one typically uses regularization schemes as, for example, Ridge regularization. In other cases, one is interested in finding sparse networks in which many of the parameters are zero. This can also be achieved by imposing an appropriate cost function on the parameters \cite{Ne20}.

In all these cases, the machine performing the classification task is defined by the form of the functions $g_\mu(\underline x)$ and the regularizing cost function is given by $H(\underline x)$ which introduces some penalty to prevent that the parameters of the network behave in a way that is not the desired one.
In this work, we will be rather generic and we will not focus on any practical application. 
We will rather focus on the characterization of the landscape of the cost function $H$ subjected to the constraints $\CC$.
In particular, we want to find the ground state of the cost function defined by
\beq
\underline x^* = \underset{\underline x: g^\mu(\underline x)\geq 0 \ \forall \mu}{\rm argmin} H(\underline x)\:.
\label{generic_opt}
\eeq
In order to have a well-posed problem, we will assume that there exists a region of the phase space of $\underline x$ such that all constraints are satisfied.  We will call this region the SAT region (the terminology comes from constraint satisfaction problems in computer science \cite{FPSUZ17}).
Examples of optimization problems as Eq.~(\ref{generic_opt}) can be easily constructed and classified according to the properties of the corresponding cost function.
We list here just a set of general cases in which the cost function is:
\begin{itemize}
\item \emph{separable}: this means that each degree of freedom tries to optimize a cost function that is independent of the configuration of the other degrees of freedom. An example in this class is 
\beq
H(\underline x) = \frac 12 \sum_{i=1}^N (x_i-1)^2
\eeq
but other non-convex potentials with local minima may be chosen.
\item \emph{non-separable and convex}: the cost function of each degree of freedom depends on the configuration of part or the whole system and the global cost is a convex function when it is considered in the whole phase space (in absence of the excluded volume constraints). As an example one can consider:
\beq
H(\underline x) = \frac 1 P\sum_{\mu=1}^P x_i \eta^\mu_i\eta^\mu_j x_j
\eeq
being $\underline \eta^\mu$ i.i.d. random vectors.
Another example is the following: consider the vector $\underline x$ to be unconstrained. Then one wants to minimize the volume of the phase space of the vector $\underline x$, for example by minimizing $|\underline x|^2$, such that there is a configuration that satisfies all constraints. This problem is related to the packing problem and has been analyzed in \cite{FP16, FPUZ15, FPSUZ17, Yo18, BIUWZ18, FHU19, IUZ19, I20}, see also \cite{GPW20} for a review with the corresponding connection to deep learning.
\item \emph{non-separable and non-convex}: we have again a non-separable cost which is also non-convex.
As an example, one can think of high dimensional random Gaussian functions \cite{CC05}
\beq
H(\underline x) = \frac 1N \sum_{i<j<k} J_{ijk}x_ix_jx_k
\eeq 
and the coupling constants are Gaussian variables of the order of $1/N$. In other words, one can superimpose a spin glass landscape on top of a continuous constraint satisfaction problem (CCSP) identified by the set of constraints.
\end{itemize}
This class of optimization problems is quite vast. In the following, we will consider the simplest case of having a simple separable convex optimization cost and the simplest set of non-convex excluded volume constraints. 
We will show that the non-convexity in the constraints brings about glassiness and marginal stability even when the cost function is convex by itself.

\section{A toy model}
To be concrete we consider a simple toy model.
We assume that each variable $x_i$ wants to minimize a simple convex cost function but it is constrained to be in a possibly non-convex region of phase space.
We consider the simplest cost function that is
\beq
H(\underline x) = \frac 12 \sum_{i=1}^N (x_i-1)^2
\label{cost_func}
\eeq
and we want to minimize it under the constraints that
\beq
	g^\mu(\underline x) \geq 0,  \ \ \ \ \ \ \forall \mu=1,\ldots, M = \alpha N
\label{ineq}
\eeq
with
\beq
g^\mu(\underline x)= \frac 1{\sqrt N} \underline \xi^\mu\cdot \underline x-\sigma
\label{perceptron}
\eeq
where $\underline \xi^\mu=\{\xi^\mu_1,\ldots, \xi^\mu_N\}$ is a random vector whose components are Gaussian random variables with zero mean and unit variance. 
Furthermore, we constrain $\underline x$ to be on the $N$-dimensional sphere $|\underline x|^2=N$.
The constraints in Eqs.~(\ref{ineq})  and (\ref{perceptron}) define the perceptron continuous constraint satisfaction problem (CCSP).
Eqs.~\eqref{cost_func}, \eqref{ineq} and \eqref{perceptron} define a quadratic programming problem with inequality constraints \cite{BV04}. However, as we will see, the topology of the phase space and the nature of the constraints are such that the global optimization problem can become non-convex.

In \cite{Ga88,DG88,FP16, FPSUZ17} the properties of the space of solutions of this CCSP have been extensively studied.
As a function of $\sigma$, one has a critical value $\alpha_J(\s)$ of the density of constraints below which the CCSP is SAT, meaning that with probability one (for $N\to \infty$) the phase space defined by the constraints has a finite positive volume, and above which the problem is UNSAT, meaning that there is no configuration of $\underline x$ that satisfies all constraints.

In \cite{FPSUZ17} the UNSAT phase of the perceptron has been studied by associating a harmonic energy cost to the unsatisfied constraints. In \cite{FP16} the SAT phase of the perceptron CCSP has been characterized. The present setting, instead, is quite different: we still consider the SAT phase of the perceptron CCSP, corresponding to $\alpha<\alpha_J(\sigma)$, but on top of it we want to optimize the cost function $H(\underline x)$ of Eq. (\ref{cost_func}). It is therefore a problem of constrained optimization.

The nature of the constraints changes completely when $\sigma$ passes from being positive, where the CCSP is convex \cite{FP16}, to a negative value, where the solution of the CCSP lies at the intersection of non-convex domains which may be in general non-convex.

In order to study the properties of the landscape of this optimization problem,
we consider the partition function defined as
\beq
Z = \int \de \underline x \exp\left[-\beta H(\underline x)\right] \left[\prod_{\mu=1}^M \theta\left(g^\mu(\underline x)\right)\right]
\eeq
where $\beta =1/T$ is the inverse of the temperature $T$ and $\theta(x)$ is the Heaviside step function.
Given the partition function, one constructs the average free energy as
\beq
{\rm f} = -\frac{1}{\b N} \overline{\ln Z}
\eeq
where the overline denotes the average over the random vectors $\xi$s.
To compute the properties of the landscape of the cost function, we need to study the behavior of the free energy in the zero-temperature limit $\beta\to \infty$.

The average of the logarithm of the partition function can be computed using the replica method \cite{MPV87}
\beq
{\rm f} = -\frac 1{\b N} \lim_{n\to 0} \partial_n \overline {Z^n}\:.
\eeq
The average over the $n$-th power of the partition function can be obtained by considering $n$ to be integer and then taking the analytic continuation down to $n\to 0$.
Performing the integral over the random vectors $\xi$s (the details are essentially very close to what is reported in \cite{FPSUZ17} and we will omit them) we obtain
\beq
\overline{Z^n} = \int \de \hat Q \exp\left[NA(\hat Q)\right]
\eeq
where $A(Q)$ is an action and $Q$ is an $(n+1)\times (n+1)$ integration matrix.
For $N\to \infty$ we can evaluate the integral by saddle point.
The expression of the function $A(\hat Q)$ is given by
\beq
A(\hat Q) = \frac 12 \ln \det \hat Q -\beta(1-m) +\alpha \ln \ZZ
\eeq
where we have assumed that $\hat Q_{1a}=\hat Q_{a1}=m$ independently on $a$\footnote{The fact that $Q_{1a}$ is independent on the replica index is a property that holds at saddle point. Indeed one can show that at the saddle point $Q_{1a} = \frac{1}{N}\sum_{i=1}^N \overline{\langle x_i\rangle}$ which is therefore independent of $a$.}.
At the saddle point we have the following interpretation for the overlap matrix $Q$:
\beq
\begin{split}
\hat Q_{ab} &= \frac 1N \underline x_a \cdot \underline x_b \ \ \ \ \ \ \ a,b> 1\\ 
\hat Q_{1a} &= \hat Q_{a1}=\frac 1N \sum_{i=1}^N \underline [\underline x_a]_i=m\ \ \ \ \ \ \ a> 1
\end{split}
\eeq
Therefore $\hat Q_{1a}=\hat Q_{a1}=m$ is an order parameter that quantifies the overlap between the configuration $\underline x_a$ and the minimum $[1,1,...,1]$ of $H(\underline x)$, 
while $\hat Q_{ab} = \frac 1N \underline x_a \cdot \underline x_b$, for $a,b>1$, is the overlap between the configurations of different replicas of the system.
Note that the spherical constraint on $\underline x$ imposes that $\hat Q_{aa}=1$.
The local partition function $\ZZ$ is given by
\beq
\ZZ =\left. \exp\left[\frac{1}{2}\sum_{a,b=1}^n Q_{ab} \frac{\partial^2}{\partial h_a\partial h_b}\right] \prod_{c=1}^n \theta\left[h_c-\sigma\right] \right|_{\underline h=0}
\eeq
where we have defined $Q$ as the matrix obtained from $\hat Q$ by removing the first row and column.  
The saddle point equations for the matrix $\hat Q$ can be obtained in full generality following the same steps as in \cite{FPSUZ17}.
Here we will not repeat the computation and we will consider the replica symmetric solution and its stability.

\subsection{The replica symmetric solution and its stability}
\begin{figure}
\centering
\includegraphics[width=\columnwidth]{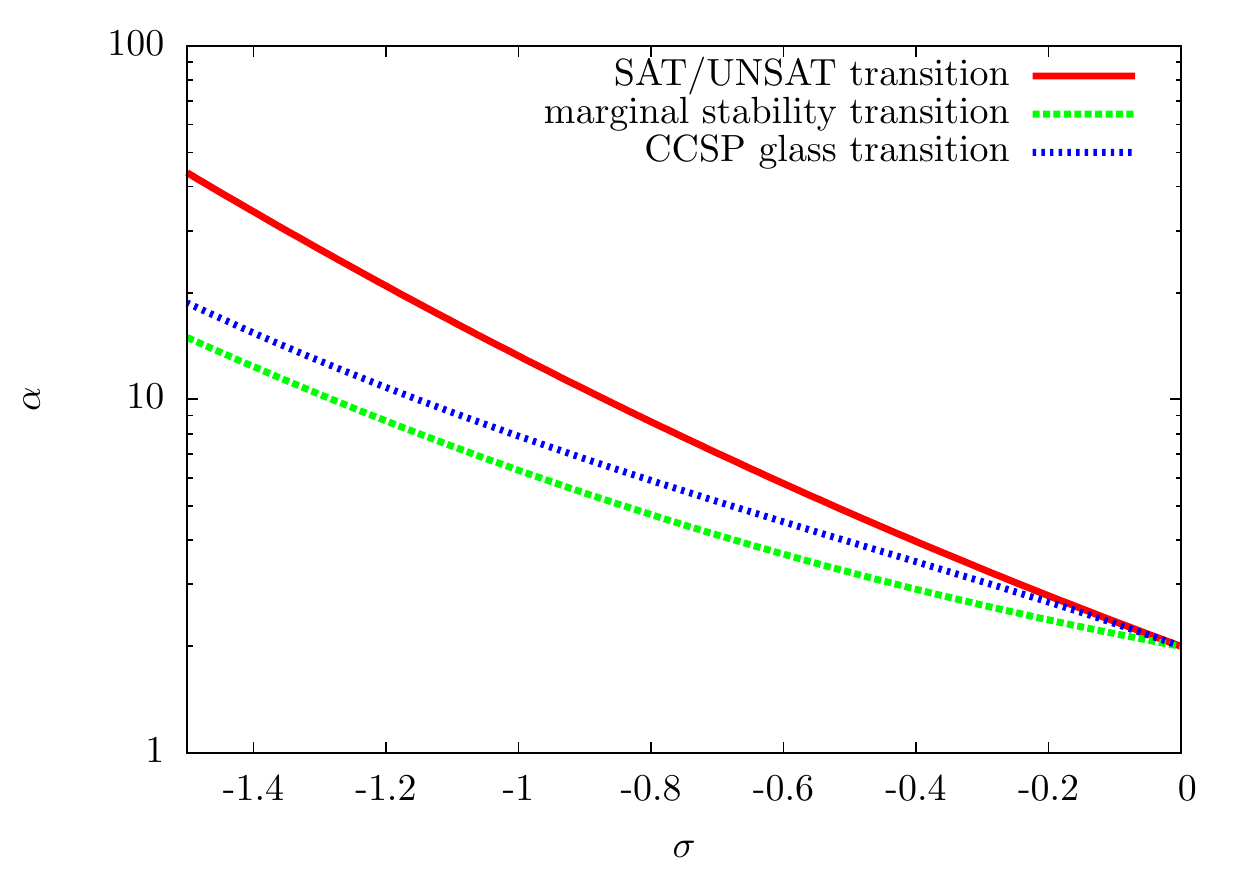}
\caption{The phase diagram of the toy optimization problem we are considering.
The model is defined only below the SAT/UNSAT transition line for which we plotted in red the replica symmetric approximation. The SAT/UNSAT transition line represents where the volume of configurations allowed by the constraints shrinks to zero. Below the green dashed line, the minimum of the cost function is unique and the replica symmetric solution is stable, meaning that the stability condition Eq. (\ref{DAT_condition}) is satisfied. The dashed green line corresponds to Eq. (\ref{DAT_condition}) becoming a strict equality. Above this line, instead, the system enters in a glassy phase characterized by multiple minima.
This phase is described by replica symmetry breaking. In the replica symmetric phase, the global minimum of the cost function is characterized by a hypostatic number of marginally satisfied constraints. Upon reaching the marginal stability line the number of those constraints becomes equal to the number of degrees of freedom so that the minimum becomes isostatic. We plot in blue also the glass transition line of the perceptron CCSP, whose equation is given in Refs. \cite{FP16, FPSUZ17} and is independent of the choice of the optimization function $H(\underline x)$ since it only depends on the perceptron constraints. Below this line, the phase space defined by the constraints is ergodic while right above ergodicity is broken and one has replica symmetry breaking.}
\label{Fig:PD}
\end{figure}

The replica symmetric ansatz to the solution of the saddle point equations amounts to have $Q_{a\neq b} = q$ for all $a\neq b$.
The order parameters $m$ and $q$ are then given by the following saddle point equations
\beq
\begin{split}
\frac{q-m^2}{(1-q)^2} &= \alpha \int\de h P(q,h) (f'(q,h))^2\\
\frac{m}{1-q}=\beta
\end{split}
\label{SP_RS}
\eeq
where
\beq
\begin{split}
f(q,h) &= \ln \g_{1-q}\star \theta(h)\\
P(q,h)&=\gamma_q(h+\s)
\end{split}
\eeq
and we used the notation according to which the prime is the derivative w.r.t. $h$, $\star$ denotes a convolution operation while
\beq
\g_A(h)= \frac{1}{\sqrt{2\pi A}} e^{-h^2/(2A)}\:.
\eeq
The stability condition of the replica symmetric solution is given by studying the Hessian $\frac{\partial^2 A(\hat Q)}{\partial\hat Q_{ab}\partial\hat Q_{cd}}$ of the function $A(\hat Q)$ \cite{AT78, MPV87, FPSUZ17}.
The replica symmetric solution is stable if the saddle point solution satisfies the inequality
\beq
\frac{1}{(1-q)^2} \geq   \alpha \int\de h P(q,h) (f''(q,h))^2\:.
\label{DAT_condition}
\eeq
When Eq.~(\ref{DAT_condition}) becomes a strict equality, the model enters in a glassy regime where multiple minima appear, and the model is characterized by replica symmetry breaking and marginal stability \cite{MPV87}.
The equations \eqref{SP_RS}-\eqref{DAT_condition} have been derived in the finite temperature regime.
However, we are interested in looking at the zero-temperature limit which corresponds to the ground state of the cost function.
In order to take the $T\to 0$  limit, we assume that in this regime
\beq
q=1-\chi T\ \ \ \ \ \ T\to 0^+
\label{zero_T}
\eeq
which gives $m=\chi$. Substituting this relation in the first equation of Eqs.~(\ref{SP_RS}) we get that $\chi$ satisfies
\beq
\frac{1-\chi^2}{\chi^2} = T^2\alpha \int \de h P(1,h) (f'(1,h))^2\:.
\eeq
Furthermore, in the zero-temperature limit, we have
\beq
 f(1,h) \simeq -\frac{h^2}{2\chi T}\theta(-h)
\eeq
which gives
\beq
1-\chi^2 =  \alpha \int \de h P(1,h) h^2 \theta(-h)\:.
\eeq
The marginal stability of the RS solution is given by
\beq
1 =  \alpha \int \de h P(1,h) \theta(-h)\:.
\eeq
This relation defines the region in which replica symmetry breaking (RSB) appears.
Note that Eq.~(\ref{zero_T}) requires $\chi$ to be positive since $q$ is constrained to be $q\leq 1$.
The phase diagram of the model is plotted in Fig.~\ref{Fig:PD} where we considered only the case of $\sigma<0$ which gives rise to non-convex excluded volume constraints.
The red line in the phase diagram represents the replica symmetric approximation for the SAT/UNSAT (jamming) transition line.
Above this line, there is no configuration of $\underline x$ that satisfies all constraints and therefore the optimization problem makes sense only in the region below this line.
This line is simply obtained by taking the $\chi\to 0$ limit of the saddle point equations. 
Indeed, above this line no physically meaningful solution should exist. Therefore this line coincides with the one obtained in \cite{FP16}. Another way to see that the $\chi\to 0$ is the right limit to get the SAT/UNSAT line is by noting that at saddle point $\chi=m$. We expect that, on this line, only one solution of the CCSP is left and, by rotational invariance of the disorder, it is orthogonal to the global minimum of the cost function. This implies that $m=0$.

The dashed green line instead corresponds to the line that separates a strictly stable phase below it, where the ground state of the cost function is unique and not critical, from a marginally stable phase where the landscape of the cost function is glassy and local minima are marginally stable. Note that the CCSP defined by the perceptron constraints also undergoes a replica symmetry breaking transition where the space of solutions disconnects before reaching the red line. However this line, plotted also in Fig.~\ref{Fig:PD}, is different from the one signaling the onset of glassiness of the cost function and, in particular, it is closer to the SAT/UNSAT transition line. Therefore we get that \emph{even if the CCSP is replica symmetric, the landscape of the cost function may be glassy.}

\subsection{The distribution of gaps}
To characterize the phase diagram, we compute the distribution of the gaps defined as the values of $g^\mu$ computed in the configuration that represents the minimum of the cost function.
Using the same approach of \cite{FPSUZ17}, we can show that the distribution of $g^\mu$ is given by
\beq
\rho(h) =\frac{1}{N}\sum_{\mu=1}^M \langle\delta(h-g^\mu(\underline x))\rangle = c\delta(h) + \rho^+(h).
\eeq
In the replica symmetric region, we have that $\rho^+(h)$ is given by
\beq
\rho^+(h)= \alpha P(1,h)\theta(h)
\eeq
and 
\beq
c=\alpha-\alpha \int_{0}^\infty \de h P(1,h) = \alpha\int_{-\infty}^0 \de P(1,h)\:.
\eeq
Therefore, when the replica symmetric solution becomes marginally stable, the system becomes isostatic, corresponding to $c\to 1$, meaning that the number of marginally satisfied constraints equals the number of degrees of freedom in the system.
At this point we may ask what happens in the glassy phase, beyond the instability line where the replica symmetric solution breaks down.
We do not attempt the RSB solution of the model but we give a simple argument.
At the jamming point we know that the allowed phase space of configurations shrinks to a point.
This point is jamming critical and isostatic \cite{FPSUZ17}.
Therefore we get that, at jamming as well as at the instability transition point, the system is isostatic. 
We, therefore, conjecture that the RSB phase is again isostatic and jamming critical.
This would imply that we have another optimization problem for which the landscape is critical everywhere far from jamming as it happens for other cases \cite{FSU19, FSU20, SU21}.

\section{Karush-Kuhn-Tucker Dynamics}
In order to solve constrained optimization problems, one can rely on 
the so called Karush-Kuhn-Tucker (KKT) conditions.
We consider a dynamical version of such conditions given by the following dynamical system
\beq
\begin{split}
\dot x_i(t) &= -\zeta(t) x_i(t) - (x_i(t)-1)\\
&+ \frac 1{\sqrt N} \sum_{\mu=1}^M \xi^{\mu}_i f_\mu(t)\\
\dot f_\mu(t) &= - f_\mu(t)g_\mu(t)
\end{split}
\label{KKT_dyn}
\eeq
where $\zeta(t)$ is a Lagrange multiplier needed to enforce the spherical constraint on the variables $x_i$. 
It is simple to show that a fixed point of these equations is a solution of the optimization problem.
Indeed if at long times $g_\mu>0$, we have $f_\mu=0$. Instead if we have $g_\mu=0$ we have $f_\mu>0$.
The variables $f_\mu$, which are the Lagrange multipliers of the KKT conditions, are nothing but the forces that the constraints are exerting in order to fix the corresponding variables $g$s to zero.
Therefore this system of equations, when it reaches a fixed point, directly gives the statistics of forces corresponding to the $g_\mu$ that are identically equal to zero.

\begin{figure}
	\centering
	\includegraphics[width=1\columnwidth]{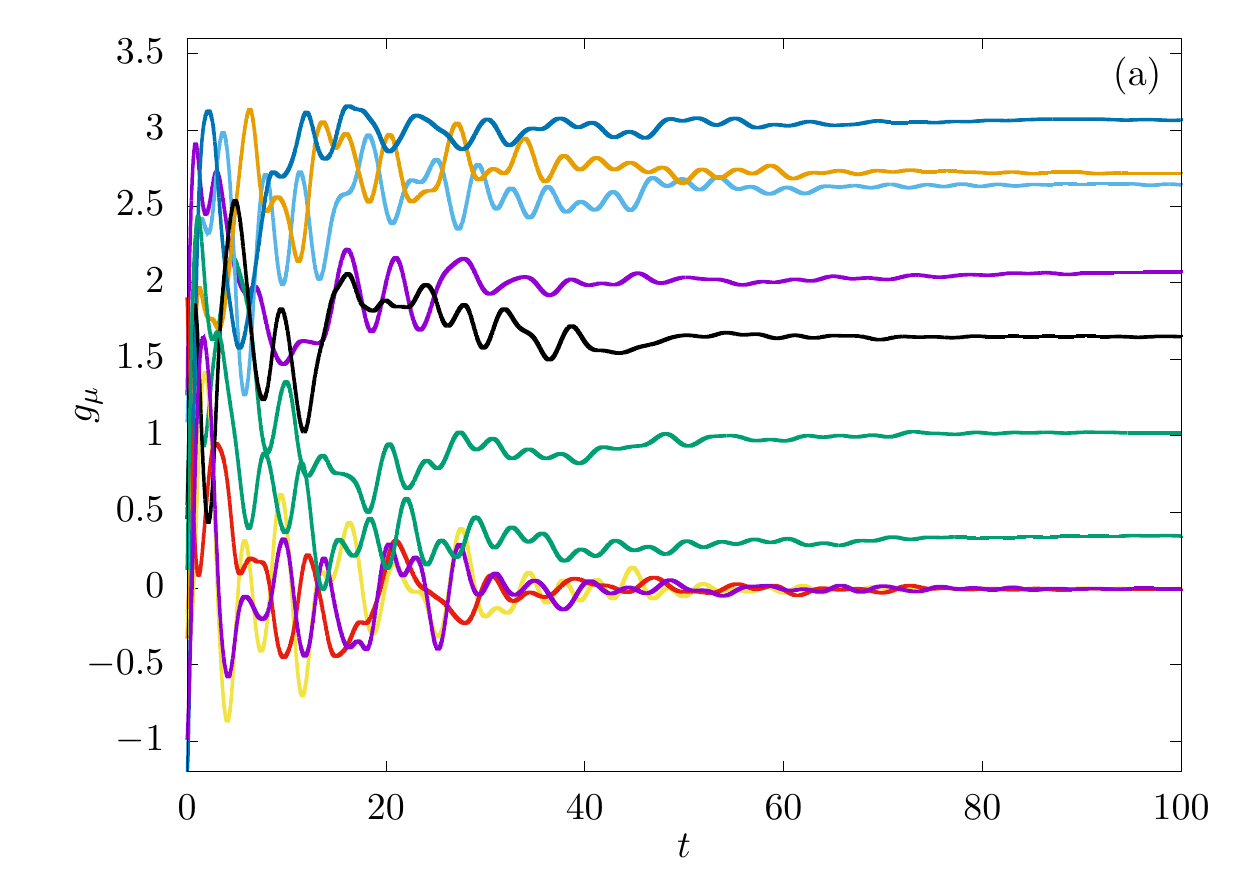}
	\includegraphics[width=1\columnwidth]{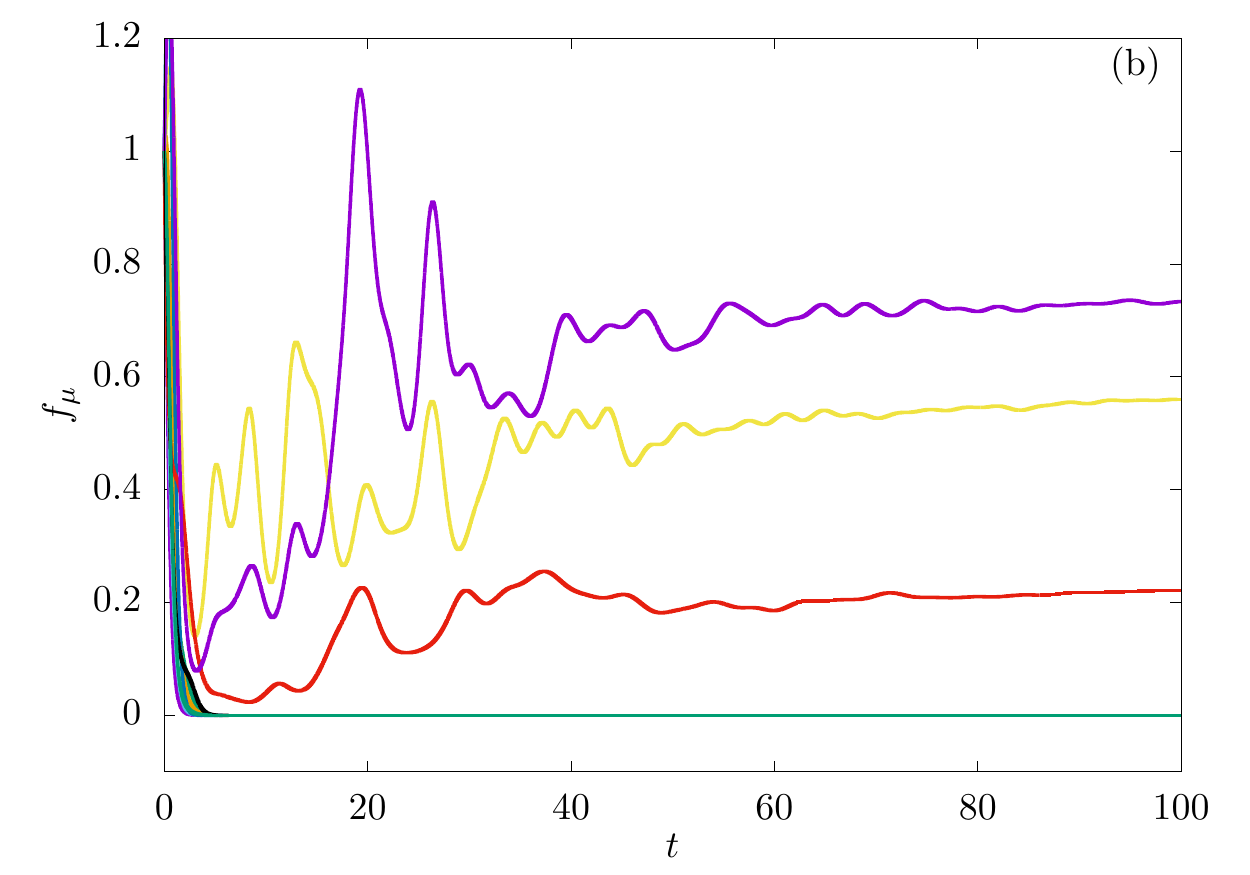}
	\caption{Time evolution of $g_{\mu}$ (panel (a)) and $f_{\mu}$ (panel (b)) in the RS phase ($\alpha=5$, $\sigma=-1$ in this example), each color corresponding to a different $\mu=1,...,10$, obtained by numerically integrating Eq. \eqref{KKT_dyn} with an explicit Runge-Kutta method of order 5 \cite{DP80-rungekutta}. 
	In the RS phase, the system quickly converges to the steady-state.
	We notice that in the steady-state of the RS phase, for a given $\mu$, $g_{\mu}> 0$ and $f_{\mu} =0 $ or $g_{\mu}= 0$ and $f_{\mu} >0 $. 
	In this experiment, the $f_{\mu}$ have been initialized to $1$ while the $x_i$ have been initialized as standard Gaussian variables.
	We used $N=100$.}
\end{figure}

This set of dynamical equations can be analyzed exactly in the $N\to \infty$ limit through dynamical mean-field theory.
We first start by considering the case in which the dynamical system is initialized at random for the $x_i$ while we will fix $f_\mu(0)=1$ which is a convenient choice.
Then, to analyze the dynamics in the $N\to \infty$ limit, we consider the Martin-Siggia-Rose-Jensenn-De-Dominicis dynamical path integral \cite{CC05} defined as
\beq
Z_{\rm dyn} = \overline{\int \DD(\underline f, \underline{\hat f})\DD(\underline x, \underline{\hat x})\DD(\underline r, \underline{\hat r}) \exp[S_{\rm dyn}]}
\eeq 
where the dynamical action is given by
\beq
\begin{split}
S_{\rm dyn} &= i\sum_{\mu=1}^M \int \de t \left\{\hat f_\mu(t) \left[\dot f_\mu(t)\right.\right.\\
&\left.\left.+f_\mu(t)\left(r_\mu(t)-\sigma\right)\right] +\hat r_\mu(t) r_\mu(t) \right\}\\
&+i \sum_{i=1}^N \int \de t \hat x_i(t) \left[\dot x_i  +\zeta(t) x_i(t)\right.\\
&\left. + (x_i(t)-1)- \frac 1{\sqrt N} \sum_{\mu=1}^M \xi^{\mu}_i f_\mu(t) \right] \\
&-\frac i{\sqrt N}\sum_{i=1}^N \int \de t \sum_{\mu=1}^M \hat r_\mu(t) \xi^\mu_i x_i(t)
\end{split}
\eeq
and the overline stands for the average over the random patterns.
Taking the average over these vectors gives
{\medmuskip=-1mu
\thinmuskip=-1mu
\thickmuskip=-1mu
\beq
\begin{split}
&\overline{\exp\left[ - \frac{i}{\sqrt N}\sum_{\mu=1}^M \sum_{i=1}^N \xi^\mu_i \int \de t \left(\hat x_i(t) f_\mu(t) + \hat r_\mu(t) x_i(t)\right)\right]}\\
&=\exp\bigg[ -\frac 12 \sum_{\mu=1}^M \int \de t \de t' \left[ f_\mu(t)f_\mu(t') D(t,t') \right.\\
&\left.+ \hat r_\mu(t)\hat r_\mu(t') C(t,t') +2i \hat r_\mu(t)f_\mu(t') R(t,t')  \right] \bigg]
\end{split}
\eeq}
where we have introduced the dynamical order parameters
\beq
\begin{split}
C(t,t') &= \frac 1N \underline x(t)\cdot \underline x(t') \\
D(t,t') &= \frac 1N \hat{\underline x}_i(t)\cdot \hat{\underline x}(t')\\
R(t,t') &= -\frac 1N \underline x(t)\cdot i\hat{\underline x}(t') 
\end{split}
\eeq
As usual, at the saddle point level $D(t,t')=0$ by causality.
The single gap effective process then gives the equation
\beq
r(t) = \int_0^t \de t' R(t,t')f(t') + \eta(t)
\label{gap_cav}
\eeq
where the statistics of the noise $\eta(t)$ is Gaussian and given by 
\beq
\overline{ \eta(t)} = 0 \ \ \ \ \ \ \overline{\eta(t)\eta(t')}= C(t,t')\:.
\eeq
Finally we have 
\beq
f(t) = \exp\left[-\int^t_0 \de s (r(s)-\s)\right]\:.
\eeq
Note that Eq.\eqref{gap_cav} must be understood as a distributional equation. Furthermore, we have used that at the initial time $t=0$ 
forces are initialized to $f_\mu(0)=1$. This initialization does not affect the endpoint of the dynamics and provides a way to enforce that all forces are either zero or positive.
The single-site effective process is instead
{\medmuskip=-1mu
\thinmuskip=-1mu
\thickmuskip=-1mu
\beq
\begin{split}
\dot x(t) &= -\zeta(t)x(t) - (x(t)-1) + \Xi(t) - \int_0^t\de s K(t,s)x(s)
\end{split}
\eeq
}
where the statistics of the Gaussian noise $\Xi(t)$ is given by
\beq
\overline{\Xi(t)}=0 \ \ \ \ \ \ \ \ \ \overline{\Xi(t)\Xi(t')} = \alpha \langle f(t)f(t')\rangle = M(t,t')
\eeq
and the kernel $K(t,s)$ is defined as
\beq
K(t,s) = \alpha\left.\frac{\delta \langle f(t)\rangle_\eta}{\delta P(s)}\right|_{P=0}
\eeq
where the perturbation $P(t)$ is an additive perturbation on the right hand side of Eq.~\eqref{gap_cav}.
Therefore we obtain the following equations for correlation, response and magnetization $m(t) = \sum_i x_i(t)/N$
{\medmuskip=-1mu
\thinmuskip=-1mu
\thickmuskip=-1mu
\beq
\begin{split}
\partial_t m(t) &= -\zeta(t)m(t) -(m(t)-1)\\
&- \int_0^t\de s K(t,s)m(s)\\
\partial_t C(t,t') &= -\zeta(t)C(t,t') - (C(t,t')-m(t')) \\
&- \int_0^t\de s K(t,s)C(t',s)+\int_0^{t'} \de s M(t,s) R(t',s)\\
\partial_t R(t,t') &= -\zeta(t) R(t,t') -R(t,t') +\delta(t-t')\\
&- \int_{t'}^t\de s K(t,s)R(s,t')\:.
\end{split}
\eeq
}
Note that the Lagrange multiplier $\zeta(t)$ can be obtained directly from the equation for $C(t,t')$.

\begin{figure}
	\centering
	\includegraphics[width=1\columnwidth]{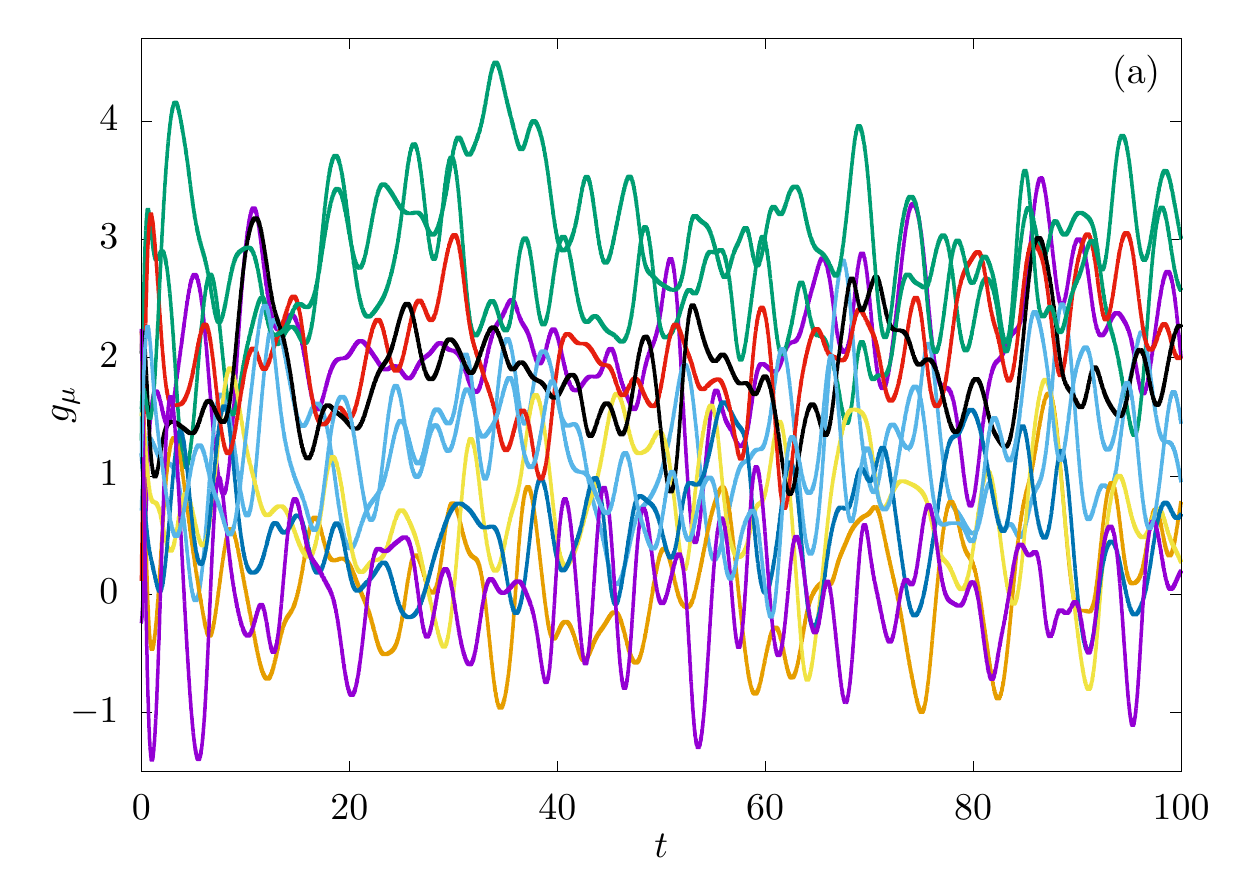}
	\includegraphics[width=1\columnwidth]{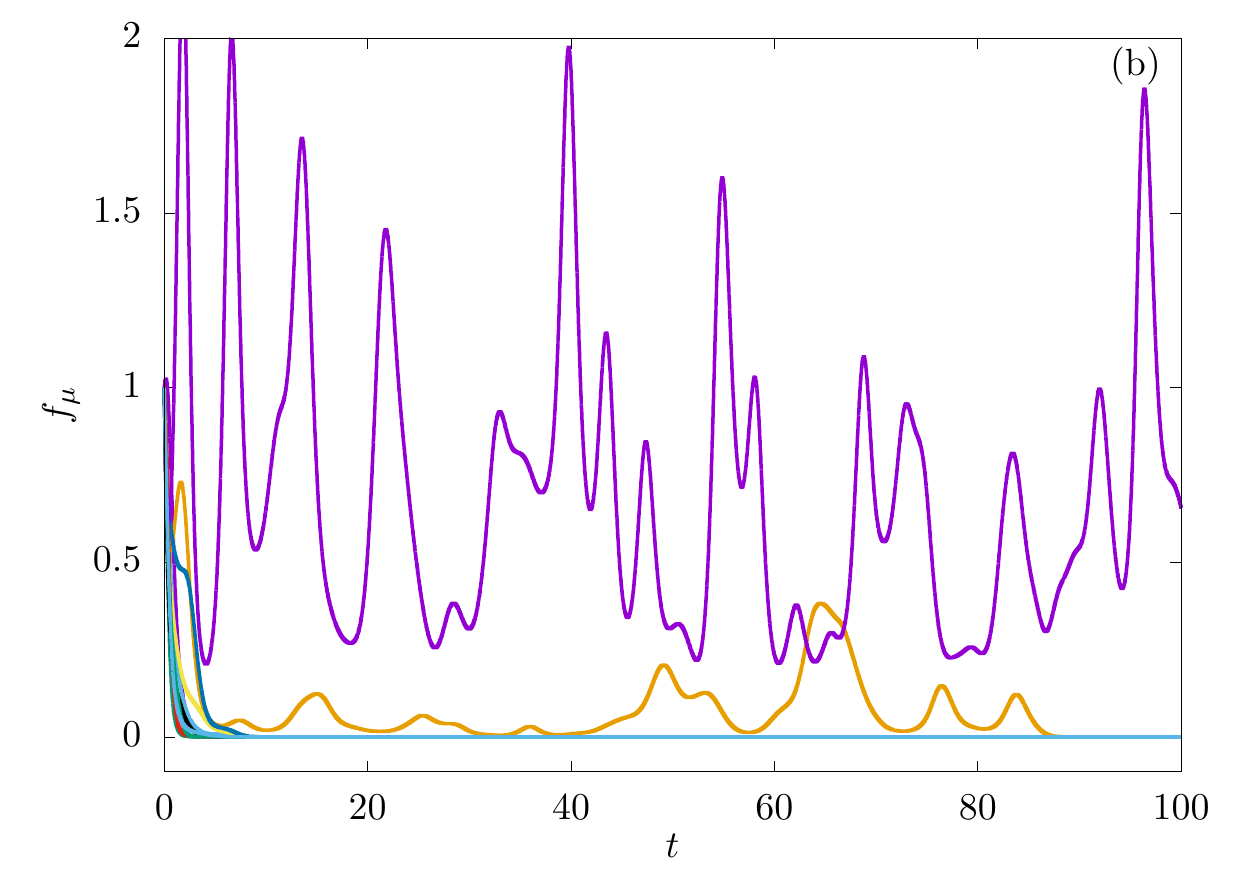}
	\caption{Time evolution of $g_{\mu}$ (panel (a)) and $f_{\mu}$ (panel (b)) in the RSB phase ($\alpha=8$, $\sigma=-1$ in this example), each color corresponding to a different $\mu=1,...,10$, obtained by numerically integrating Eq. \eqref{KKT_dyn} with an explicit Runge-Kutta method of order 5 \cite{DP80-rungekutta}. 
	Since the RSB phase is marginally stable, the time evolution is chaotic and does not converge to a stationary state.
	In this experiment, the $f_{\mu}$ have been initialized to $1$ while the $x_i$ have been initialized as standard Gaussian variables.
	We used $N=100$.}
\end{figure}

From these dynamical equations it follows that
\beq
m(t) = \int_0^t \de s R(t,s)\:.
\eeq
which is a relation connecting the response function $R(t,s)$ to the magnetization $m(t)$.
\subsection{The replica symmetric solution}
Let us consider what happens in the region where replica symmetry is unbroken.
If we define
\beq
\chi = \lim_{t\to \infty}\int_0^t \de s R(t,s)
\eeq
we immediately have
\beq
m_\infty\equiv \lim_{t\to \infty}m(t)=\chi\:.
\eeq
Considering Eq.\eqref{gap_cav} and taking the $t\to\infty$ limit we get
that
\beq
r_\infty = \chi f_\infty +\eta
\eeq
being $\eta$ a normal Gaussian random variable.
This equation tells us that
if $f_\infty=0$ then $r_\infty$ is a Gaussian random variable constrained to be such that $r_\infty-\sigma$ is positive.
Therefore the gap distribution is
\beq
P(h) =\gamma_1(h+\sigma) \:.
\eeq 
On the other hand, if $r_\infty=\sigma$ then $f_\infty>0$ and its distribution coincides with the one of the random variable $(\sigma-\eta)/\chi$ with the constraint that $\sigma-\eta>0$.
Finally we need to establish an equation for $\chi$ and show that it coincides with the one coming from the replica approach.
We consider the long time limit of the equation for $m(t)$ which gives
\beq
0=-\z_\infty \chi -(\chi-1) -K_\infty \chi
\label{KK_eq}
\eeq
where 
\beq
K_\infty = \lim_{t\to \infty}\int_0^\infty \de s K(t,s) \:.
\eeq
On the other hand the long time limit of the equation for $M$ is obtained by considering the long time limit of the equation for $C$. This gives
\beq
0=-\z_{\infty} +1-\chi -K_{\infty}+M_{\infty}\chi
\eeq
which can be combined with Eq.\eqref{KK_eq} to get
\beq
M_{\infty} \chi^2  = 1-\chi^2\:.
\label{M1}
\eeq
From the definition of $M_\infty$ we finally have
\beq
M_\infty = \alpha \langle f_\infty^2\rangle = \alpha \int\frac{\de \eta}{\sqrt{2\pi}}e^{-\eta^2/2} \frac{(\eta+\s)^2}{\chi^2}\theta(\eta+\sigma)
\label{M2}
\eeq
and therefore using Eqs.~(\ref{M1}) and (\ref{M2}) we get Eq.~\eqref{SP_RS}.
This concludes the derivation of the replica symmetric equation from the KKT dynamics.

We implemented numerically the algorithm of Eq.~\eqref{KKT_dyn} and we found that in the RS phase the algorithm goes very quickly to the solution of the optimization problem. 
In the RSB phase, instead, it seems that the algorithm does not converge and therefore we believe that in this region it makes a chaotic surf on the isostatic landscape.

\section{Conclusions}
We have considered the generic setting of optimization 
problems under non-convex excluded volume constraints.
We have analyzed a simple example of this kind in which the cost function is separable and convex
and in which the non-convex excluded volume constraints are modeled by a negative perceptron.

We have found that, when the number of constraints is low enough, the cost function has a simple ground state 
which is captured by the replica symmetric solution of the model.
At high density of constraints, instead, the optimization landscape undergoes an RSB transition where minima become marginally stable.
Remarkably, we find that the RSB transition point happens \emph{before} the point in which the accessible phase space defined by the constraints splits into glassy regions. 
We have also shown how to recover these results from the dynamical mean-field theory of the KKT algorithm.
We leave the analysis of more complex cost functions for future work.

\acknowledgments
This work was supported by ''Investissements d'Avenir'' LabExPALM (ANR-10-LABX-0039-PALM).  A.S. thanks LPTMS where part of this work has been done and the Simons Foundation Grant (No. 454941 Silvio Franz and No. 454953 Matthieu Wyart) for support.

\bibliography{HS}

\end{document}